\documentclass[10pt,twocolumn,oneside]{IEEEtran}
\usepackage{cite}
\usepackage{array}
\usepackage{changepage}

\ifCLASSINFOpdf
   \usepackage[pdftex]{graphicx}
\else
\fi
\usepackage{bm}
\usepackage{graphicx}
\usepackage{subfigure}

\usepackage{multirow}
\usepackage{multicol}

\usepackage[cmex10]{amsmath}

\usepackage{algorithm} 
\usepackage{algorithmic} 
\newcommand*\varhrulefill[1][0.4pt]{\leavevmode\hrule height#1\kern0pt}

\usepackage{enumerate}

\usepackage{changepage}

\usepackage{mdwmath}
\usepackage{makecell,rotating,multirow,diagbox}

\usepackage{multirow}
\usepackage{booktabs}
\usepackage{dcolumn}
\usepackage{amssymb}

\usepackage{url}

\usepackage{subfloat}
\usepackage{color}

\usepackage{flushend}
\usepackage{fancyhdr}
\fancypagestyle{plain}{
    \fancyhf{}
    \fancyfoot[C]{\textsc{978-1-5090-0690-8/16/\$31.00 \copyright 2016 IEEE}}
    
    }
    \fancypagestyle{empty}{
    \fancyhf{}
    \fancyfoot[C]{\textsc{}}
    
    }

\begin{document}
\title{A-Optimal Sampling and Robust Reconstruction for Graph Signals via Truncated Neumann Series}
\author{\IEEEauthorblockN{Fen Wang, Yongchao Wang, \emph{Member, IEEE}, and {Gene Cheung, \emph{Senior Member, IEEE}}}\\
\renewcommand{\baselinestretch}{1.0}
\thanks{The research of F. Wang and Y. Wang was supported in part by National Science Foundation of
China under grant 61771356, 111 project of China under grant B08038.}
\thanks{F. Wang and Y. Wang are with State Key Lab. of ISN, School
of Telecom. Engineering, Xidian University, No.2 Taibai South Road,
Xi'an, 710071, Shaanxi, China.(Email:ychwang@mail.xidian.edu.cn).}
\thanks{{G. Cheung is with National Institute of Informatics, 2-1-2, Hitotsubashi, Chiyoda-ku, Tokyo, 101-8430, Japan. (Email:cheung@nii.ac.jp).}}
}
\maketitle
\thispagestyle{empty} 

\vspace{0.01cm}
\begin{abstract}
Graph signal processing (GSP) studies signals that live on irregular data kernels described by graphs.
One fundamental problem in GSP is sampling---from which subset of graph nodes to collect samples in order to reconstruct a bandlimited graph signal in high fidelity.
In this paper, we seek a sampling strategy that minimizes the mean square error (MSE) of the reconstructed bandlimited graph signals assuming an independent and identically distributed (iid) noise model---leading naturally to the A-optimal design criterion.
To avoid matrix inversion, we first prove that the inverse of the information matrix in the A-optimal criterion is equivalent to a Neumann matrix series.
We then transform the truncated Neumann series {based sampling problem} into an equivalent expression that replaces eigenvectors of the Laplacian operator with a sub-matrix of an ideal low-pass graph filter.
Finally, we approximate the ideal filter using a Chebyshev matrix polynomial.
We design a greedy algorithm to iteratively minimize the simplified objective.
For signal reconstruction, we propose an accompanied signal reconstruction strategy that reuses the approximated filter sub-matrix and {is provably more robust than conventional least square recovery}.
Simulation results show that our sampling strategy outperforms two previous strategies in MSE performance at comparable complexity.
\end{abstract}

\vspace{-0.2in}
\section{Introduction}
\vspace{0cm}
\textit{Graph signal processing} (GSP) is the study of discrete signals that live on irregular data kernels described by graphs \cite{emerging,gsp}.
One fundamental problem in GSP is {\textit{sample selection}}---optimally select a subset of graph nodes from which to collect samples such that an assumed bandlimited signal can be reconstructed in high fidelity\footnote{{Sampling can also be done via \textit{aggregation} \cite{aggregation}: observe the same signal after different graph shifts but only at one node.}}{\cite{sampling,randomSampling,Uncertainty,sketching}}.
Under noiseless conditions, \cite{largedata} proved that a \emph{qualified sampling set} that leads to perfect signal reconstruction requires only full column rank of a sampling matrix, and {empirically showed that it can be accomplished with high probability via random node selection for a connected graph via large experiments.}
Random node selection can result in a poor condition number in the reconstruction matrix, however.
In response, authors in \cite{towards,semisupervised,AO} proposed an efficient sampling strategy based on spectral proxies that selects stable nodes for unique reconstruction without full eigen-decomposition.
If the observed samples are corrupted by noise, \cite{SCsampling} {adopts} an E-optimality criterion for sampling, which minimizes the worst case reconstruction error.

However, most graph sampling works\footnote{{\cite{Uncertainty} proposed a generic greedy procedure, {termed as minimum Frobenius norm (MFN) based selection,} to minimize the A-optimality criterion but no fast algorithms. We compare against \cite{Uncertainty} in our experiments.}} do not adopt and optimize a minimum mean square error (MMSE) objective directly, which leads naturally to an A-optimality design criterion assuming an independent and identically distributed (iid) additive noise model \cite{Boyd}.
Beyond the fact that graph sampling is inherently combinatorial in nature, one main difficulty lies in the computation of the inverse of an information matrix, which in general has complexity $\mathcal{O}(n^3)$.
While \cite{greedybound} showed that greedy methods optimizing the A-optimality criterion can have near-optimal performance, no efficient implementation was proposed, which in general requires eigen-decomposition and/or inversion of large matrices.

In this letter, we propose a computation-efficient graph sampling strategy that addresses the A-optimality criterion directly.
We first prove that the inverse information matrix in the A-optimality criterion is equivalent to a Neumann matrix series.
We next transform the truncated Neumann series {based sampling problem} into an equivalent expression that replaces eigenvectors of the Laplacian operator with a sub-matrix of an ideal graph low-pass filter \cite{emerging,semisupervised}.
Finally, we approximate the low-pass filter with a Chebyshev matrix polynomial \cite{wavelet}---\textit{arriving at a simplified proxy that approximates the A-optimality objective but requires neither full eigen-decomposition nor matrix inversion}.
We propose a greedy algorithm to minimize the simplified objective.
For signal reconstruction, we design an accompanied reconstruction strategy that is provably more ``robust" \footnote{{By ``robustness'', we mean resilience of an estimator's presumably good MSE performance for small noise as noise variance increases.}} to large noise than the least square (LS) solution while reusing the earlier computed approximate low-pass filter sub-matrix.
Experimental results show that our A-optimality based sampling strategy outperforms previous schemes in MSE at comparable complexity.

\section{Signal Processing on Graphs}

Denote by $\mathcal{G}=(\mathcal{V},\mathcal{E},\mathbf{W})$ a graph containing a set of nodes indexed by $\mathcal{V}=\{ 1,2,...,n\}$.
$\mathcal{E}$ is the set of weighted edges.
An edge weight $\mathbf{W}(i,j)=w_{ij}$ reflects the similarity between nodes $i$ and $j$.
In this letter, we focus on connected, undirected graphs with no multiple edges and adopt the symmetric normalized Laplacian matrix $\pmb{\mathcal{L} }=\mathbf{I}-{\mathbf{D}^{ - 1/2}}\mathbf{W}{\mathbf{D}^{ - 1/2}}$ as the variation operator, where ${\mathbf{D} = \textrm{diag}\{ {d_1},...,{d_n}\} }$ and ${{d_i} = \sum_j {{w_{ij}}}}$.
Assuming that the eigen-decomposition of $\pmb{\mathcal{L} }$ is $\pmb{{\mathcal{L}} }= \mathbf{V} \mathbf{\Sigma} {{\mathbf{V}^{T}}} $ where $\mathbf{\Sigma}=\textrm{diag}\{\lambda _i\}$ with ${{\lambda _1} \le ... \le {\lambda _n}}$, $\mathbf{V}={\{ {\mathbf{v}_1}, ... ,{\mathbf{v}_n}\} }$ and $\mathbf{V}\mathbf{V}^T=\mathbf{I}$, then the \textit{graph Fourier transform} (GFT) of a graph signal $\mathbf{x} \in {\mathbb{R}^{n}}$ is defined as $\tilde {\mathbf{x}} = {\mathbf{V}^T}\mathbf{x}$ and the inverse GFT is $\mathbf{x} = \mathbf{V}\tilde {\mathbf{x}}$.
A graph signal is called \textit{bandlimited} if there exists a number $K\in\mathcal{V}$ such that ${{{\tilde {x}}_i} = 0},\textrm{for} ~\textrm{all}~ {i > K}$ \cite{SCsampling}.
The smallest such ${K}$ is called the \textit{bandwidth} of ${\mathbf{x}}$.
Graph signals with bandwidth at most ${K}$ are called $K$-bandlimited ($K$-BL) graph signals, which are expressed as
${\mathbf{x} = {\mathbf{V}_K}{{\tilde {\mathbf{x}}}_K}}$.
${\mathbf{V}_K}$ means the first ${K}$ columns of ${\mathbf{V}}$ and ${{{\tilde {\mathbf{x}}}_K}}$ denotes the first ${K}$ elements of ${\tilde {\mathbf{x}}}$.

\emph{\textbf{Definition 1}}\cite{codeguide}:~{In order to select $m$ elements from $\mathbf{x}$ to produce ${\mathbf{x}_{\mathcal{S}}=\mathbf{Cx} \in {{\rm{\mathbb{R}}}^m}}$ with ${\left| \mathcal{S} \right| = m}$ and ${\mathcal{S} \subseteq \mathcal{V}}$, we define the \textit{sampling matrix} $\mathbf{C} \in \mathbb{F}^{m \times n}$ as}
\begin{equation}\label{sampling}
\begin{split}
{{\mathbf{C}_{ij}} = \left\{ \begin{array}{l}
1,~~~{j = {\mathcal{S}_i}};\\
0,~~~\textrm{otherwise},
\end{array} \right.}
\end{split}
\end{equation}
where $\mathcal{S}$ is the set of sampling indices, $\mathcal{S}_i$ means the $i$-th element of set $\mathcal{S}$, and ${\left| \mathcal{S} \right|}$ is the number of elements in ${\mathcal{S}}$.

$\mathcal{S}^{c}$ denotes the complement set of $\mathcal{S}$.
$|\mathcal{E}|$ is the number of edges.
For any matrix ${\rm{\mathbf{A}}}$, we adopt the notation ${{\mathbf{A}_{{\mathcal{S}_1}{\mathcal{S}_2}}}}$ to denote the sub-matrix of ${\mathbf{A}}$ with rows indexed by ${\mathcal{S}_1}$ and columns indexed by ${\mathcal{S}_2}$.
${{\mathbf{A}_{{\mathcal{S}}{\mathcal{S}}}}}$ is simplified to ${{\mathbf{A}_{{\mathcal{S}}}}}$.
$\mathbf{I}$ is the identity matrix whose dimension depends on the context.

\section{Sample Selection for Noisy $K$-BL Graph Signals}

A sampled $K$-BL graph signal can now be written as ${{\mathbf{x}_\mathcal{S}}= \mathbf{C} {\mathbf{V}_K}{{\tilde {\mathbf{x}}}_K}}$.
In noiseless condition, $\mathbf{x}$ can be perfectly recovered from ${\mathbf{x}_\mathcal{S}}$ when ${\textrm{rank}(\mathbf{C} {\mathbf{V}_K}) = K}$, using the LS solution \cite{SCsampling}:
\begin{equation}\label{reconstructed signal}
\begin{split}
{\bar{\mathbf{x}}}= {\mathbf{V}_K}{\left( {\mathbf{C} {\mathbf{V}_K}} \right)^{\dagger}}\mathbf{x}_\mathcal{S},
\end{split}
\end{equation}
where ${ (\cdot)^ {\dag} }$ means the pseudo-inverse operator.

When corrupted by noise, a sampled $K$-BL graph signal is $\mathbf{y}_\mathcal{S}=\mathbf{x}_\mathcal{S}+\mathbf{n}$.
Using \eqref{reconstructed signal} as the recovery method, we get an estimator $\hat{\mathbf{x}} = {\mathbf{V}_K}{\left( {\mathbf{C} {\mathbf{V}_K}} \right)^{\dag}}(\mathbf{x}_\mathcal{S}+\mathbf{n})$.
Assuming that noise $\mathbf{n}$ is iid with zero mean and unit variance, the covariance matrix of the reconstruction error is
${C_{{{\hat{\mathbf{x}}}}}}={\mathbf{V}_K}\left[ ({\mathbf{C}{\mathbf{V}_K})^{T}}\mathbf{C} {\mathbf{V}_K} \right]^{ - 1}{\mathbf{V}^{T}_K}$ \cite{AO}.
By the theory of optimal experiments design \cite{experiments}, minimizing the trace of the covariance matrix leads to the known \textit{A-optimality} criterion:
\begin{equation}\label{original formulation}
\begin{split}
\mathbf{C}^{*}=\mathop {\arg \min }\limits_{\mathbf{C} \in \mathbb{F}^{m\times n}}\textrm{tr}\left({[ {{{(\mathbf{C}{\mathbf{V}_K})}^{T}}\mathbf{C}{\mathbf{V}_K}} ]^{ - 1}}\right).
\end{split}
\end{equation}
{Notice that the A-optimality criterion coincides with the MMSE criterion.}

An alternative criterion is to minimize the largest eigenvalue of $C_{\hat{\mathbf{x}}}$---as adopted in \cite{SCsampling}---known as the \textit{E-optimality} criterion, which minimizes the worst case signal reconstruction.

Unlike \cite{SCsampling}, we address directly the MMSE criterion in (\ref{original formulation}) to achieve smallest signal reconstruction error on average, but doing so without any matrix inversion or eigen-decomposition, where complexity is $\mathcal{O}(n^3)$ in general.

\hspace{0cm}
\subsection{Matrix Inversion Approximation}

\emph{{\textbf{Proposition 1:}}} The inverse matrix in \eqref{original formulation} exists and is equivalent to its Neumann series, \textit{i.e.},
\begin{equation}\label{Neumann series}
\begin{split}
[ {{{(\mathbf{C}{\mathbf{V}_K})}^{T}}\mathbf{C}{\mathbf{V}_K}} ]^{-1}={\sum_{l = 0}^\infty {{{[ {\mathbf{I} - {{(\mathbf{C}{\mathbf{V}_K})}^{T}}\mathbf{C}{\mathbf{V}_K}} ]}^l}} },
 \end{split}
\end{equation}
if the selected matrix ${{ \mathbf{C}{\mathbf{V}_K}}}$ is full column rank, \textit{i.e.},
\begin{equation}\label{full rank}
\begin{split}
\textrm{rank}{{{\left( \mathbf{C}{\mathbf{V}_K} \right)}}} = K.
\end{split}
\end{equation}

\emph{\textbf{Proof:}} For simplicity, we denote $\mathbf{\Psi}  = {\left( {\mathbf{C}{\mathbf{V}_K}} \right)^{T}}\mathbf{C}{\mathbf{V}_K}$, $\mathbf{\Phi}=\mathbf{I}-\mathbf{\Psi}$ and let ${\delta _1} \le \ldots  \le {\delta _K}$ be the eigenvalues of $\mathbf{\Phi}$.
The \emph{Neumann series theorem} [18, Section 5.6] states that if $ \rho(\mathbf{\Phi})=\mathop {\max }\limits_i \left| {{\delta _i}} \right| <1$, then the Neumann series $\mathbf{I} + \mathbf{\Phi} + {\mathbf{\Phi}^2} +  \cdots $ will converge to ${(\mathbf{I} - \mathbf{\Phi})^{ - 1}}$, which exactly implies \eqref{Neumann series}.
From the definition of $\mathbf{C}$, ${\mathbf{C}^{T}}\mathbf{C} = {\left[ \begin{array}{l}
 {\mathbf{I}_\mathcal{S}}~~~{\bf{0}} \\
 {\bf{0}}~~~~{\kern 1pt}{\bf{0}} \\
 \end{array} \right]}$ under appropriate permutation.
Hence, $\forall \mathbf{x} \in {\mathbb{R}^K}$ and $\left\| \mathbf{x} \right\|_2 = 1$,
\begin{equation}\label{product}
\begin{split}
\begin{array}{l}
 {\mathbf{x}^{T}}\mathbf{\Psi} \mathbf{x} = {\left( {{\mathbf{V}_K}\mathbf{x}} \right)^{T}}\left( {{\mathbf{C}^{T}}\mathbf{C}} \right)\left( {{\mathbf{V}_K}\mathbf{x}} \right) \\
~~~~~~~~= {\mathbf{b}^{T}}\left[ \begin{array}{l}
 {\mathbf{I}_\mathcal{S}}~~~{\bf{0}} \\
 {\bf{0}}~~~~ {\kern 1pt} {\bf{0}} \\
 \end{array} \right]\mathbf{b}={\mathbf{b}^{T}_{{\mathcal{S}}}}{\mathbf{b}_{{\mathcal{S}}}}, \\
 \end{array}
 \end{split}
\end{equation}
{where $\mathbf{b}={{\mathbf{V}_K}\mathbf{x}}$}.

Since ${\mathbf{b}^{T}}\mathbf{b} = {\left( {{\mathbf{V}_K}\mathbf{x}} \right)^{T}}\left( {{\mathbf{V}_K}\mathbf{x}}\right)=1$, $0 \le {\mathbf{x}^{T}}\mathbf{\Psi} \mathbf{x} \le 1$.
Because $\textrm{rank}{{{\left( \mathbf{C}{\mathbf{V}_K} \right)}}} = K$, $\mathbf{\Psi}$ is positive definite, which results in $0 < {\mathbf{x}^{T}}\mathbf{\Psi} \mathbf{x} \le 1$ and $0 \le {\mathbf{x}^{T}}\mathbf{\Phi} \mathbf{x} < 1$.
Based on the \emph{Rayleigh quotient theorem} \cite{neumannseries}, $0 \le {\delta _i}<1$ and $ \rho(\mathbf{\Phi})<1$.
$\Box$

{As discussed in \cite{SCsampling}, \eqref{full rank} is the definition of \emph{qualified sampling operators}, and it is ensured with high probability via random node selection when $m\geq K$.
The necessary condition for a \emph{qualified sampling operator} is thus $m\geq K$, which is the focus region of all sampling strategies.}
In the following analysis, {we assume that \eqref{full rank} is satisfied}.

We propose the following sampling method by substituting the inverse matrix in \eqref{original formulation} with a truncated Neumann series.
\begin{equation}\label{reformulated problem}
\begin{split}
{\mathbf{C}^*} = \mathop {\arg \min }\limits_{\mathbf{C} \in \mathbb{F}^{m\times n}} \textrm{tr}\left( {\sum\limits_{l = 0}^L {{{[ {\mathbf{I} - {{(\mathbf{C}{\mathbf{V}_K})}^{T}}\mathbf{C}{\mathbf{V}_K}} ]}^l}} } \right),
 \end{split}
\end{equation}
where $L$ is a truncation parameter.

\textit{\textbf{Proposition 2:}}
When $\mathbf{\Psi}^{-1}$ is approximated by its truncated Neumann series, the estimation error between \eqref{original formulation} and \eqref{reformulated problem} is
\begin{align}\label{trace error}
\left| {{\rm{tr}}\left[ {\sum\limits_{l = 0}^\infty  {{{ {\bf{\Phi }} }^l}} } \right] - {\rm{tr}}\left[ {\sum\limits_{l = 0}^L {{{ {\bf{\Phi }} }^l}} } \right]} \right|
= \sum\limits_{i = 1}^K {\frac{{{\delta ^{L + 1}_i}}}{{1 - {\delta _i}}}}.
\end{align}

See Appendix A in the supporting document for the proof of this result.
{Equation \eqref{trace error} implies that the proper design of $L$ depends on the sizes of $\delta_i$ which are influenced by the sampling strategy, sampling size $m$ and $\mathbf{V}_K$.}
Larger $L$ would result in a smaller truncation error.
In this letter, we set $L=10$ and show {that} its estimate error is reasonably small in Section V.

Leveraging on a property of the trace operator, we further transform \eqref{reformulated problem} into an equivalent problem {that involves an ideal low-pass graph filter $\mathbf{T}$ with cutoff frequency $\lambda_K$}.

\emph{\textbf{Theorem 1:}}  The sampling problem \eqref{reformulated problem} is equivalent to
\begin{equation}\label{simplified problem}
\begin{split}
{\mathcal{S}^*} = \mathop {\arg \min }_{\mathcal{S}:|\mathcal{S}|=m} \textrm{tr}\left[ {\sum_{l = 0}^L {{{( {\mathbf{I}}_{\mathcal{S}} - {\mathbf{T}_\mathcal{S}} )}^l}} } \right],
 \end{split}
\end{equation}
where $\mathbf{T} = {\mathbf{V}_K}{\mathbf{V}^{T}_K}\in \mathbb{R}^{n\times n}$ is an ideal low-pass graph filter (implementation to be discussed in details) and the relationship between $\mathbf{C}$ and $\mathcal{S}$ is defined in \eqref{sampling}.

\emph{\textbf{Proof:}} Denote by $\mathbf{P}={ {\mathbf{C}}^{T}}\mathbf{C}{\mathbf{V}_K}{\mathbf{V}^{T}_K}$.
 Because $\textrm{tr}(\mathbf{AB})=\textrm{tr}(\mathbf{BA})$ and $\mathbf{\Psi}  = { {\mathbf{V}^{T}_K}{\mathbf{C}^{T}} }\mathbf{C}{\mathbf{V}_K}$, $\textrm{tr}(\mathbf{\Psi}^{l})=\textrm{tr}({\mathbf{V}^{T}_K}{ {\mathbf{C}}^{T}}\mathbf{C}{\mathbf{V}_K}...\\{\mathbf{V}^{T}_K}{ {\mathbf{C}}^{T}}\mathbf{C}{\mathbf{V}_K})=\textrm{tr}({ {\mathbf{C}}^{T}}\mathbf{C}{\mathbf{V}_K}...{\mathbf{V}^{T}_K}{{\mathbf{C}}^{T}}\mathbf{C}{\mathbf{V}_K}{\mathbf{V}^{T}_K})
 =\textrm{tr}(\mathbf{P}^{l})$.
As a result,
\begin{align}\label{first transform}
&\hspace{0.3cm}\textrm{tr}(\mathbf{I}-\mathbf{\Psi})^{l}=\textrm{tr} [\mathbf{I}+\sum\limits_{d = 1}^{l}\tbinom{l}{d}(-\mathbf{\Psi})^{d}]\\
&\hspace{0cm}\mathop =\limits^{\bigtriangleup}\textrm{tr}[\mathbf{I}+\sum\limits_{d = 1}^{l}\tbinom{l}{d}(-\mathbf{P})^{d}]-n+K=\textrm{tr}(\mathbf{I}-\mathbf{P})^{l}-n+K,\nonumber
\end{align}
where $\tbinom{l}{d}$ means the binomial coefficients and $\mathop =\limits^{\bigtriangleup}$ is derived from the property of trace operation.

Since ${\mathbf{C}^{T}}\mathbf{C} = {\left[ \begin{array}{l}
 {\mathbf{I}_\mathcal{S}}~~~ {\bf{0}} \\
 {\bf{0}}~~~~{\kern 1pt} {\bf{0}} \\
 \end{array} \right]}$ under appropriate permutation, $\mathbf{P} = \left[ \begin{array}{l}
{\mathbf{T}_\mathcal{SV}} \\
~\mathbf{0} \\
 \end{array} \right]$.
 Hence, $
\mathbf{I} - \mathbf{P} = \left[ \begin{array}{l}
{\mathbf{I}}_{\mathcal{S}}- {\mathbf{T}_\mathcal{S}}~~-{\mathbf{T}_{\mathcal{S}{\mathcal{S}^c}}} \\
~~~~\mathbf{0}~~~~~~~~~~~{\mathbf{I}_{{\mathcal{S}^c}}} \\
 \end{array} \right]
 $,
which will lead to
\begin{equation}\label{structure}
\begin{split}
(\mathbf{I} - \mathbf{P})^{l} = \left[ \begin{array}{l}
({\mathbf{I}}_{\mathcal{S}}- {\mathbf{T}_\mathcal{S}})^{l}~~ \bullet\\
~~~~~~ \mathbf{0}~~~~~~~~{\mathbf{I}_{{\mathcal{S}^c}}}
\end{array} \right],
\end{split}
\end{equation}
where ``$\bullet$'' denotes a nonzero matrix whose dimension is $m\times (n-m)$.

Therefore, $\textrm{tr}(\mathbf{I} - \mathbf{P})^{l}=\textrm{tr}({\mathbf{I}}_{\mathcal{S}}- {\mathbf{T}_\mathcal{S}})^{l}+n-m$.
Combined with \eqref{first transform}, $\textrm{tr}(\mathbf{I}-\mathbf{\Psi})^{l}=\textrm{tr}({\mathbf{I}}_{\mathcal{S}} - {\mathbf{T}_\mathcal{S}})^{l}-m+K$ where $m$ and $K$ are constant during sampling, which implies Theorem 1.
$\Box$

\begin{table}\label{table1}
 \centering
\caption{Outline of The Proposed MIA Sampling Algorithm}
 \setlength{\tabcolsep}{3pt}
\begin{tabular}{p{1cm}p{7cm}}
\hline
\hline
\textbf{Input:}&$\pmb{\mathcal{L}}$, {bandwidth $K$}, sample size ${m}$ and truncation parameter $L$\\
\textbf{Output:}& Sampling set ${\mathcal{S}}$\\
Step 1.& $\mathcal{S}=\{\varnothing\}$, compute $\lambda_K$ of $\pmb{\mathcal{L}}$\\
Step 2.& {Calculate the truncated Chebyshev polynomial approximation of $h({\lambda})$ and then compute $\mathbf{T}^{\textrm{Poly}}=\sum_{j=0}^p \beta_j \pmb{\mathcal{{L}}}^j$}\\
Step 3.& \textbf{While} $\left| \mathcal{S} \right| < m$\\
~&~~~~~${\mathbf{\Gamma} _i} =\mathbf{I}_{\mathcal{S}+\{i\}}-  {{\mathbf{T}}}^{\textrm{Poly}} _{\mathcal{S} + \left\{ i \right\}}$~$\forall {i \in \mathcal{S}^{c}}$\\
~&~~~~~$u = \mathop {\arg \min }\limits_{i \in \mathcal{S}^{c}} \textrm{tr}\left(\sum\limits_{l = 0}^L {{\mathbf{\Gamma}^l_i}}\right) $\\
~&~~~~~${\mathcal{S} \leftarrow \mathcal{S} + \left\{ u \right\}}$\\
~& \textbf{end}\\
Step 4.& Return ${\mathcal{S}}$ and $\tilde{\mathbf{\Gamma}}={\sum\limits_{l = 0}^L {{{( {\mathbf{I}}_\mathcal{S} - {\mathbf{T}^{\textrm{Poly}}_\mathcal{S}} )}^l}} }$\\
\hline
\hline
\end{tabular}
\end{table}

\vspace{-0.05in}
\subsection{Chebyshev Approximation of Low-pass Filter $\mathbf{T}$}
The low-pass graph filter $\mathbf{T}$ has a kernel function $h({\lambda}) = \left\{ \begin{array}{l}
1,~~~\lambda\leq \lambda_K  \\
0,~~~\lambda> \lambda_K \\
 \end{array} \right.
$.
{We approximate this spectral kernel function via Chebyshev polynomial approximation \cite{wavelet}, after which} $\mathbf{T}$ can be expressed by a matrix polynomial in terms of $\pmb{\mathcal{L}}$, \textit{i.e.}, $\mathbf{T}^{\textrm{Poly}}=\sum _{i=1}^{n}\left(  \sum_{j=0}^p \beta_j\lambda^j_i\right)\mathbf{v}_i\mathbf{v}^{T}_i=\sum_{j=0}^p \beta_j \pmb{\mathcal{{L}}}^j$ \cite{ILSR}.
Finally, the sampling problem is formulated as
\begin{equation}\label{final problem}
\begin{split}
{\mathcal{S}^*} = \mathop {\arg \min }_{\mathcal{S}:|\mathcal{S}|=m} \textrm{tr}\left[ {\sum_{l = 0}^L {{{( {\mathbf{I}}_{\mathcal{S}} - {\mathbf{T}^{\textrm{Poly}}_\mathcal{S}} )}^l}} } \right].
 \end{split}
\end{equation}

If we solve \eqref{reformulated problem} directly, the eigenvector matrix $\mathbf{V}_K$ of dimension $n \times K$ is required.
After transforming \eqref{reformulated problem} to \eqref{final problem}, the required information is only $\lambda_K$.
{We compute $\lambda_K$ as follows.
We first compute an $M$-by-$M$ tridiagonal matrix using the {Lanczos} algorithm, which preserves $\lambda_K$ of $\pmb{\mathcal{{L}}}$ \cite{golub}.
For {a} large sparse graph, the complexity of the Lanczos algorithm is $\mathcal{O}(Mn)$.
We then compute the eigenvalues of the tridiagonal matrix using a fast multipole method, whose complexity is $\mathcal{O}(M \textrm{log} M)$ \cite{Coakley}. Hence, the combined complexity for computing $\lambda_K$ is $\mathcal{O}(Mn)$, where $K < M \ll n$.}
Along with $\lambda_K$, the computation of $\mathbf{T}^{\textrm{Poly}}$ {may be done only once during preparation}, and implemented efficiently due to the sparsity of $\pmb{\mathcal{L}}$.

\vspace{-0.08in}
\subsection{Complexity Analysis}

\vspace{-0.02in}
Optimizing the proposed criterion \eqref{final problem} is still combinatorial, so we adopt a greedy algorithm to obtain its solution, which we call the \textit{matrix inversion approximation} (MIA) sampling algorithm.
Details of the algorithm are presented in Table I.
The complexity for computing $\lambda_K$ is $\mathcal{O}(Mn)$.
Calculating the coefficients $\pmb{\beta}$ and $\mathbf{T}^{\textrm{Poly}}=\sum_{j=0}^p \beta_j \pmb{\mathcal{{L}}}^j$ has complexity $\mathcal{O}(pn)$ and $\mathcal{O}(pn|\mathcal{E}|)$ \cite{wavelet}.
Hence, in the preparation step, the complexity of MIA is $\mathcal{O}(pn|\mathcal{E}|)$.
In each sampling step, the algorithm involves matrix multiplication, where ${\mathbf{I}}_{\mathcal{S}}-\mathbf{T}^{\textrm{Poly}}_\mathcal{S} \in \mathbb{R}^{|\mathcal{S}|\times|\mathcal{S}|}$ and $|\mathcal{S}|$ gradually increases until $m$, having an asymptotic complexity of $\mathcal{O}(m^{2.373})$ \cite{matrix}.
Considering the impact from $L$ and $n$, the complexity of each search step is $\mathcal{O}(nLm^{2.373})$.
Since finally $|\mathcal{S}|=m$, the whole complexity of the sampling step is $\mathcal{O}(nLm^{3.373})$.
\begin{figure*}[htbp]
    \centering
     \subfigure[Reconstruction MSE in \textit{\textbf{G1}} at 10dB]{
    \begin{minipage}{4.3cm}
    \centering
        \includegraphics[width=1.8in,height=2.0in]{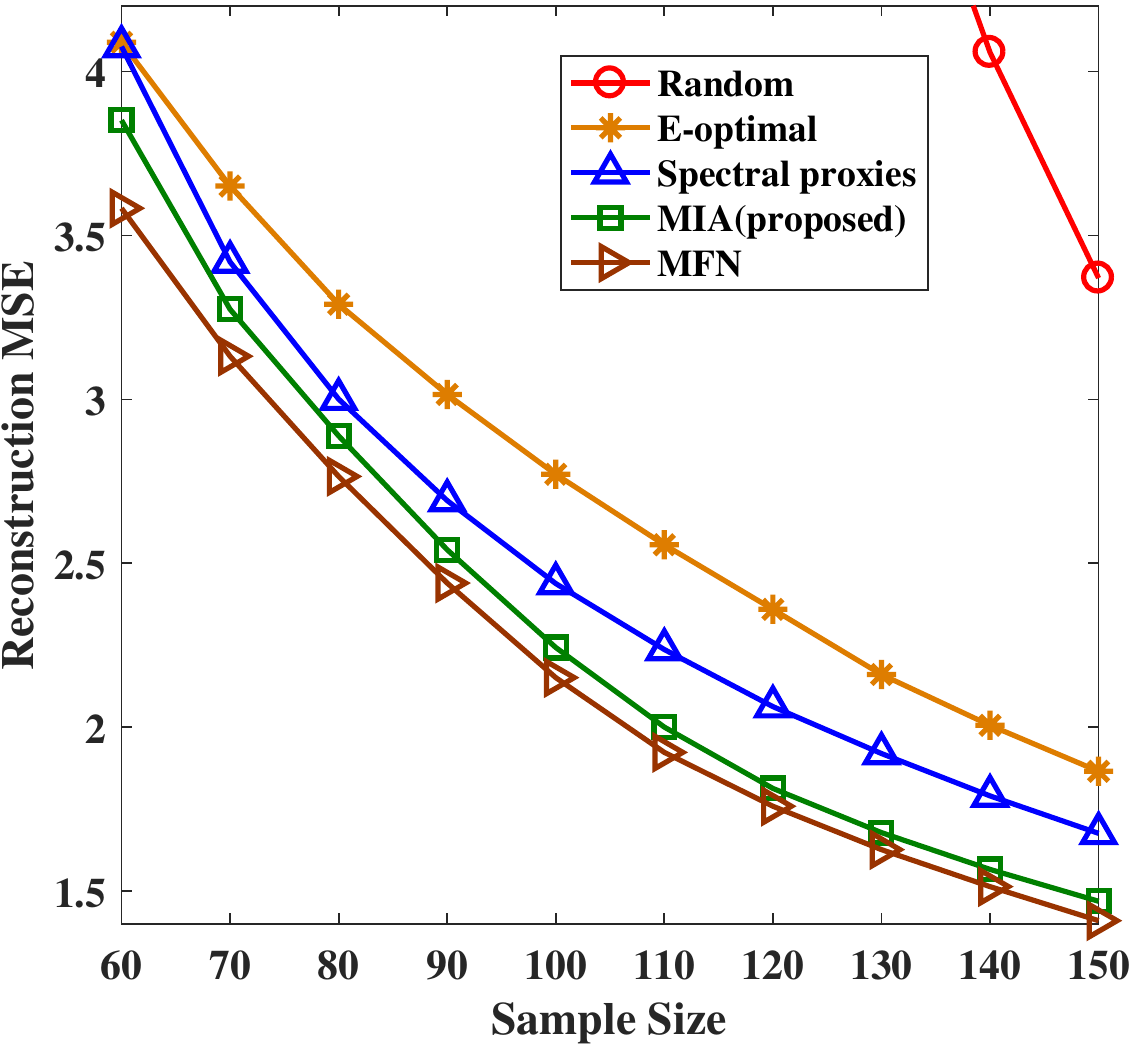}
    \end{minipage}%
    }
     \subfigure[Reconstruction MSE in \textit{\textbf{G1}} at 0dB]{
    \begin{minipage}{4.3cm}
    \centering
        \includegraphics[width=1.8in,height=2.0in]{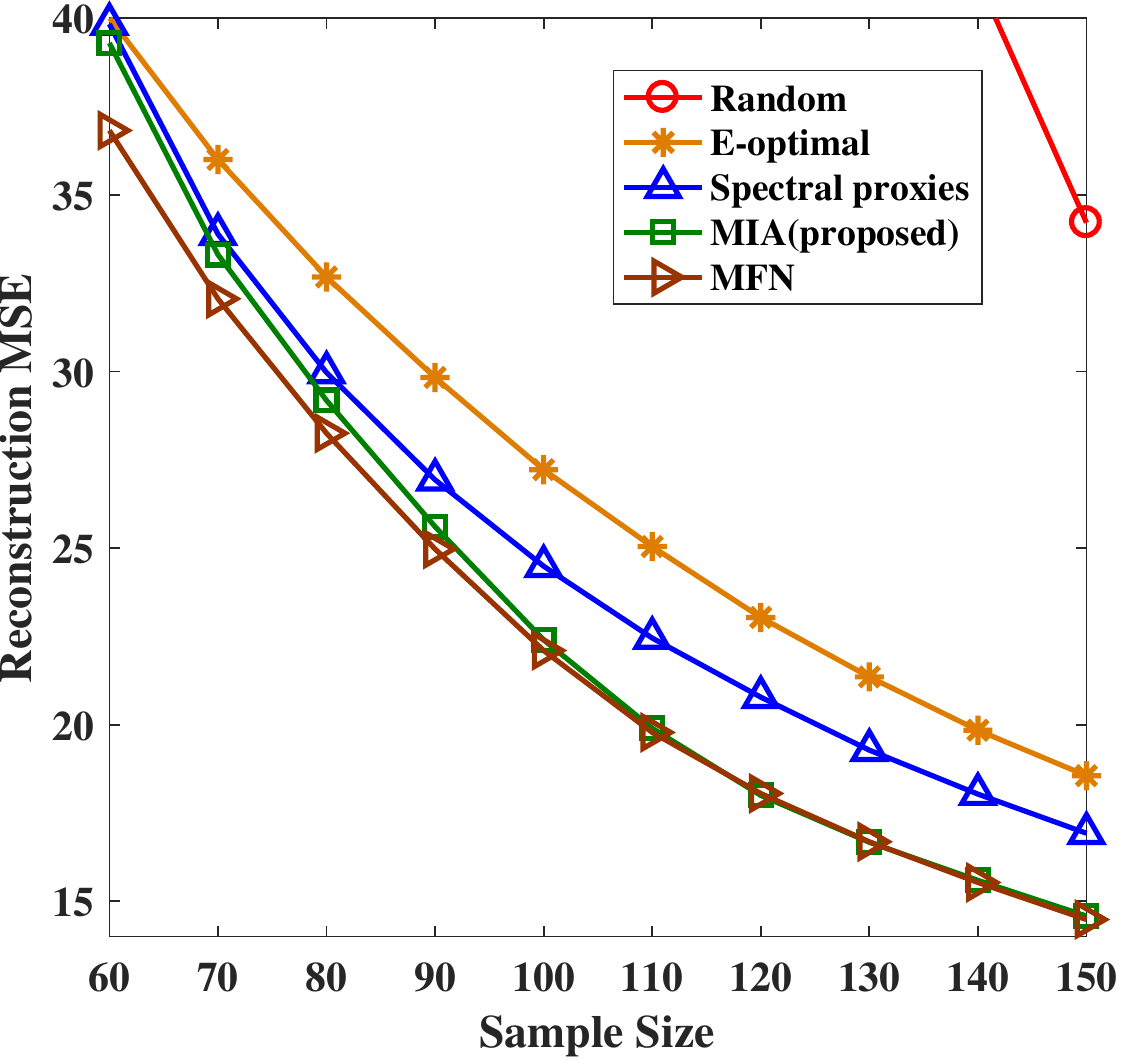}
    \end{minipage}%
    }
    \subfigure[Reconstruction MSE in \textit{\textbf{G2}} at 0dB]{
    \begin{minipage}{4.3cm}
    \centering
        \includegraphics[width=1.8in,height=2.0in]{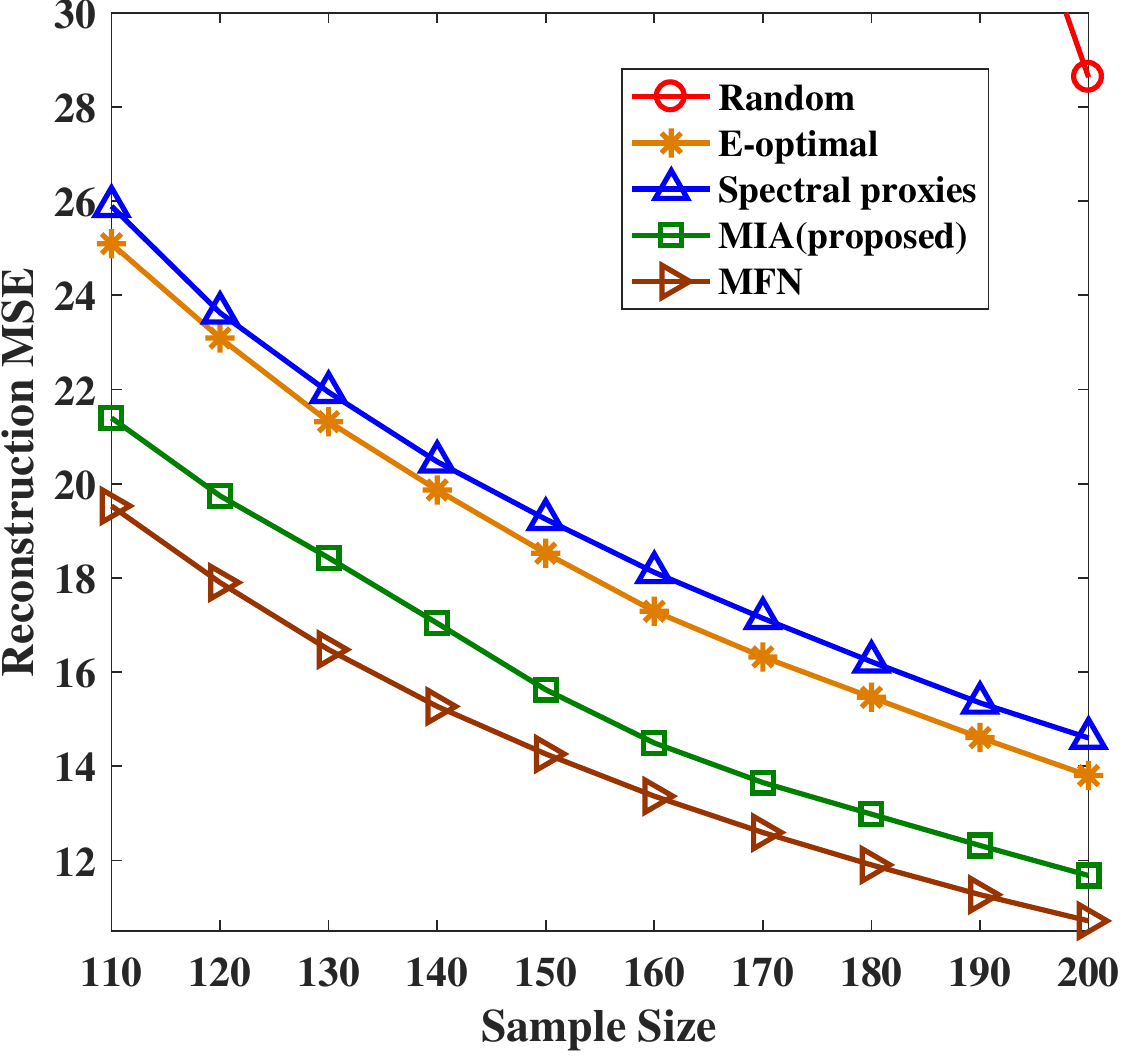}
    \end{minipage}%
    }
    \centering
    \subfigure[Complexity comparison]{
    \begin{minipage}{4.3cm}
    \centering
        \includegraphics[width=1.8in,height=2.0in]{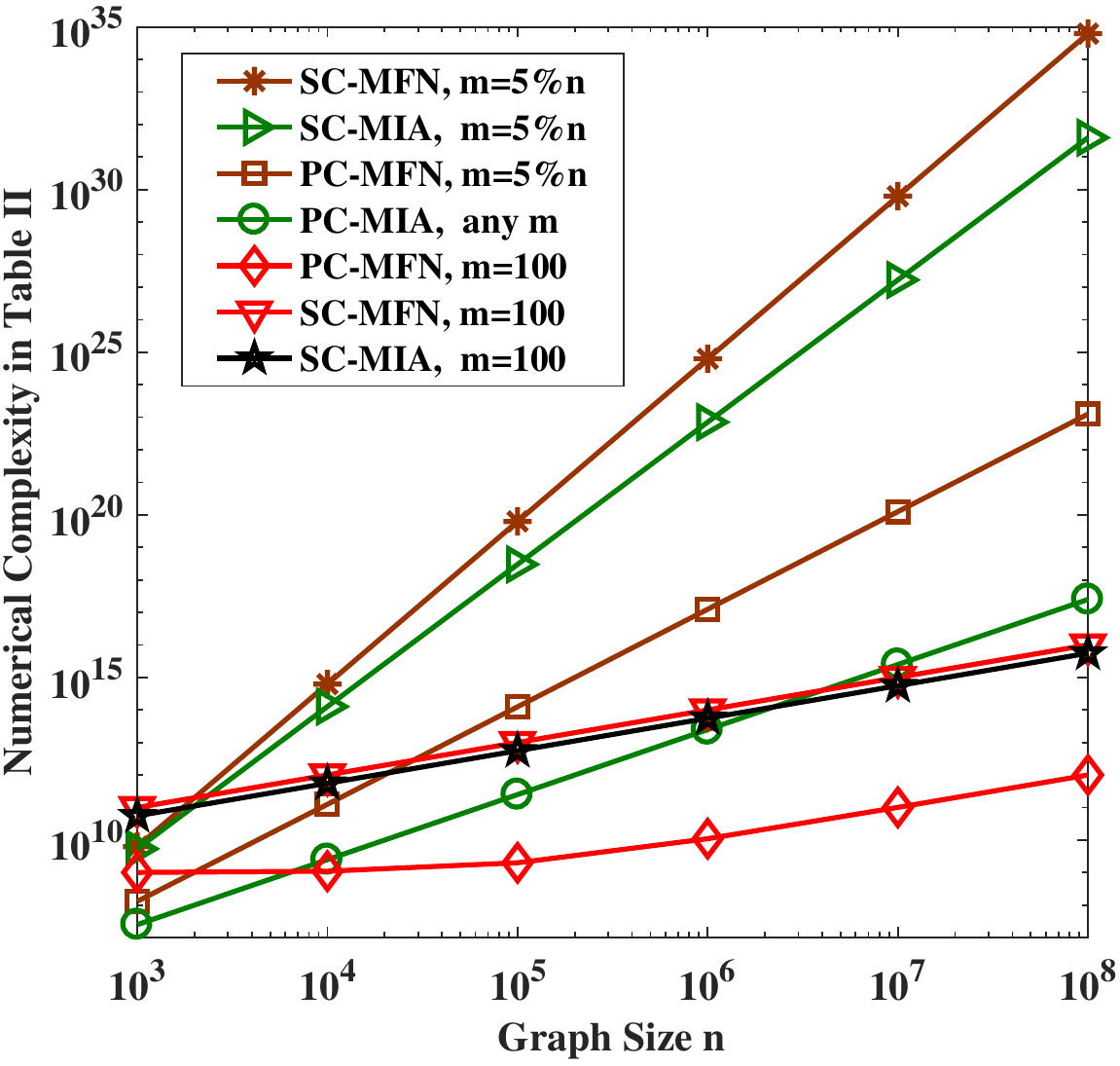}
    \end{minipage}%
    }
\caption{(a) (b) (c) Simulation results for different sampling algorithms where graph signals are all recovered by the LS reconstruction, {(d) Numerical comparison of complexity between the MFN and the MIA algorithms where ``PC'' and ``SC'' are the abbreviation of the complexity of the preparation step and that of the sampling step respectively. }}
\end{figure*}

\begin{table}
\caption{Complexity Comparison of Different Sampling Strategies}
\begin{center}
\begin{tabular}{|c|c|c|}
 \hline
\hline
  \textbf{}&\textbf{Preparation}&\textbf{Selection step}\\\hline
Spectral Proxies &\emph{NONE}&$\mathcal{O}\left({k\left| \mathcal{E} \right|m{T_2}\left( k \right)}+nm\right)$ \\\hline
E-optimal &$\mathcal{O}\left( {\left( {\left| \mathcal{E} \right|m + R{m^3}} \right){T_1}} \right)$&$\mathcal{O}\left(nm^{4}\right)$\\\hline
{MFN} &{$\mathcal{O}\left( {\left( {\left| \mathcal{E} \right|m + R{m^3}} \right){T_1}} \right)$}&{$\mathcal{O}\left(nm^{4}\right)$}\\\hline
MIA&$\mathcal{O}(pn|\mathcal{E}|)$&$\mathcal{O}\left(nLm^{3.373}\right)$\\\hline
 \end{tabular}
 \end{center}
\end{table}

Table II compares the computational complexity among different sampling strategies, in which we assume $m=K$ and adopt some results in \cite{AO}.
In the preparation step, the spectral proxies algorithm utilizes $\pmb{\mathcal{L}}$ directly, while $\mathbf{V}_K$ is necessary for {the E-optimal and the MFN algorithms.}
In the selection step, {the E-optimal and the MFN algorithms} need singular value decomposition and the first eigen-pair of $\left((\pmb{\mathcal{{L}}}^{T})^k\pmb{\mathcal{{L}}}^k\right)_{\mathcal{S}^c}$ is required for the spectral proxies algorithm.

\vspace{-0.08in}
\section{Accompanied Reconstruction Strategy}

\vspace{-0.02in}
Assuming $\textrm{rank}{{{\left( \mathbf{C}{\mathbf{V}_K} \right)}}} = K$, then according to Proposition 1, the LS solution of a graph signal is equivalent to
\begin{align}\label{Neumann recovery}
&\hspace{0cm}\hat{\mathbf{x}}={\mathbf{V}_K}\left[ ({\mathbf{C}{\mathbf{V}_K})^{T}}\mathbf{C}{\mathbf{V}_K} \right]^{-1}({\mathbf{C}\mathbf{V}_K})^{T}\mathbf{y}_\mathcal{S}\nonumber\\
&\hspace{0.25cm}={\mathbf{V}_K} {\sum\limits_{l = 0}^\infty {{{\left[ {\mathbf{I} - \mathbf{\Psi}} \right]}^l}} }{\mathbf{V}^{T}_K}\mathbf{C}^{T} \mathbf{y}_\mathcal{S}.
 \end{align}

It is easy to derive that
 \begin{align}\label{series}
&\hspace{0cm}{\mathbf{V}_K}{{{( {\mathbf{I} -\mathbf{\Psi} } )}^l}} {\mathbf{V}^{T}_K}={\mathbf{V}_K}\left[ {\sum\limits_{d = 0}^{l} \tbinom{l}{d}\mathbf{I}^{l-d}(-1)^{d} \mathbf{\Psi}^{d} }\right]{\mathbf{V}^{T}_K}\nonumber\\
&\hspace{1cm}={\sum\limits_{d = 0}^{l} \tbinom{l}{d}\mathbf{I}^{l-d}(-1)^{{d}} \left({\mathbf{V}_K} \mathbf{\Psi}^{d}{\mathbf{V}^{T}_K}\right) }\\
&\hspace{1cm}\mathop  = \limits^{\bigtriangleup}{\sum\limits_{d = 0}^{l} \tbinom{l}{d}\mathbf{I}^{l-d}(-1)^{{d}} \left({\mathbf{V}_K}{\mathbf{V}^{T}_K} \mathbf{P}^{d}\right) }=\mathbf{T}{{{({\mathbf{I} -\mathbf{P} })}^l}},\nonumber
 \end{align}where $\mathop=\limits^{\bigtriangleup}$ holds since
${\mathbf{V}_K}({\mathbf{V}^{T}_K}{ {\mathbf{C}}^{T}}\mathbf{C}{\mathbf{V}_K})...({\mathbf{V}^{T}_K}{ {\mathbf{C}}^{T}}\mathbf{C}{\mathbf{V}_K}){\mathbf{V}^{T}_K}\\={\mathbf{V}_K}{\mathbf{V}^{T}_K}\mathbf{P}^{d}$.
\begin{figure}
\begin{center}
\includegraphics[width=120pt,height=135pt]{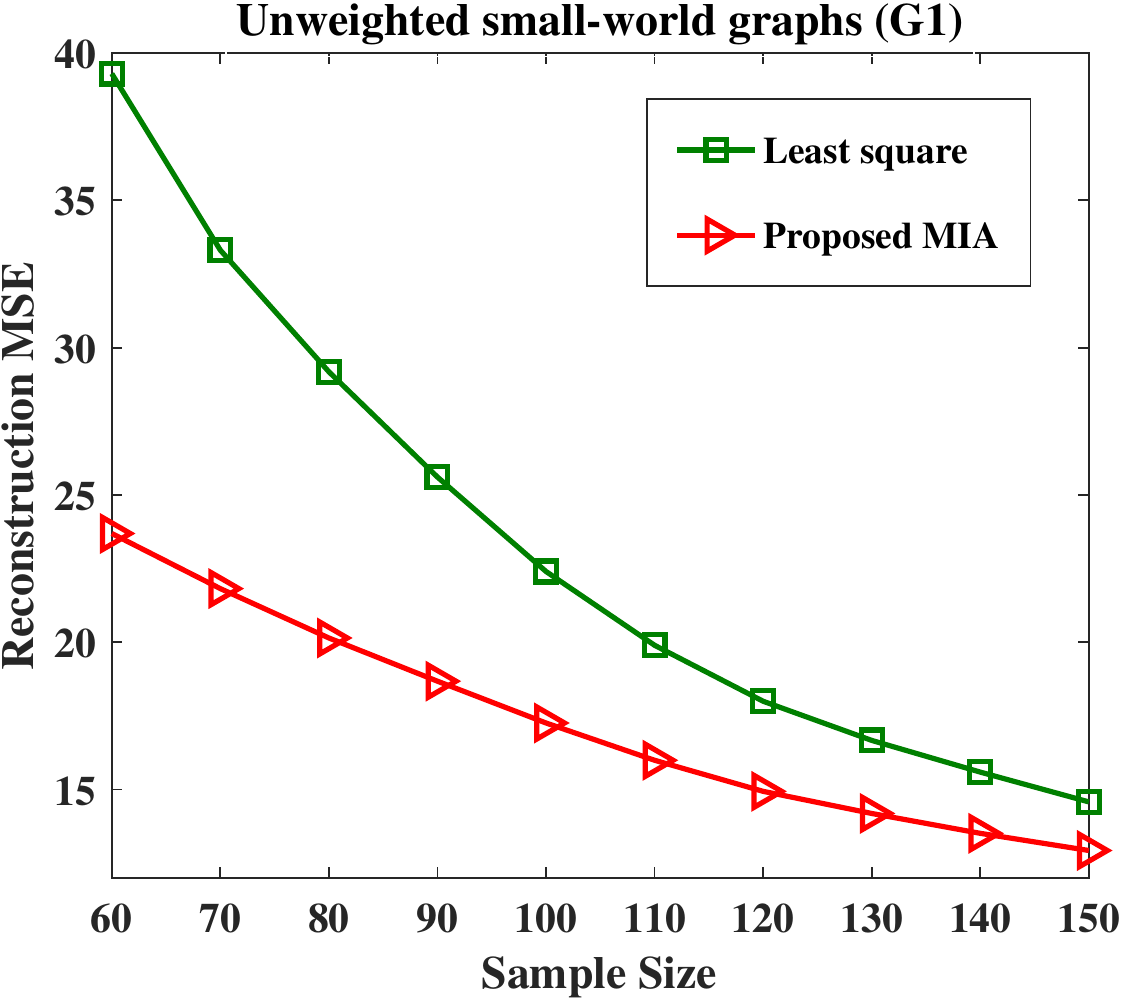}
\includegraphics[width=120pt,height=135pt]{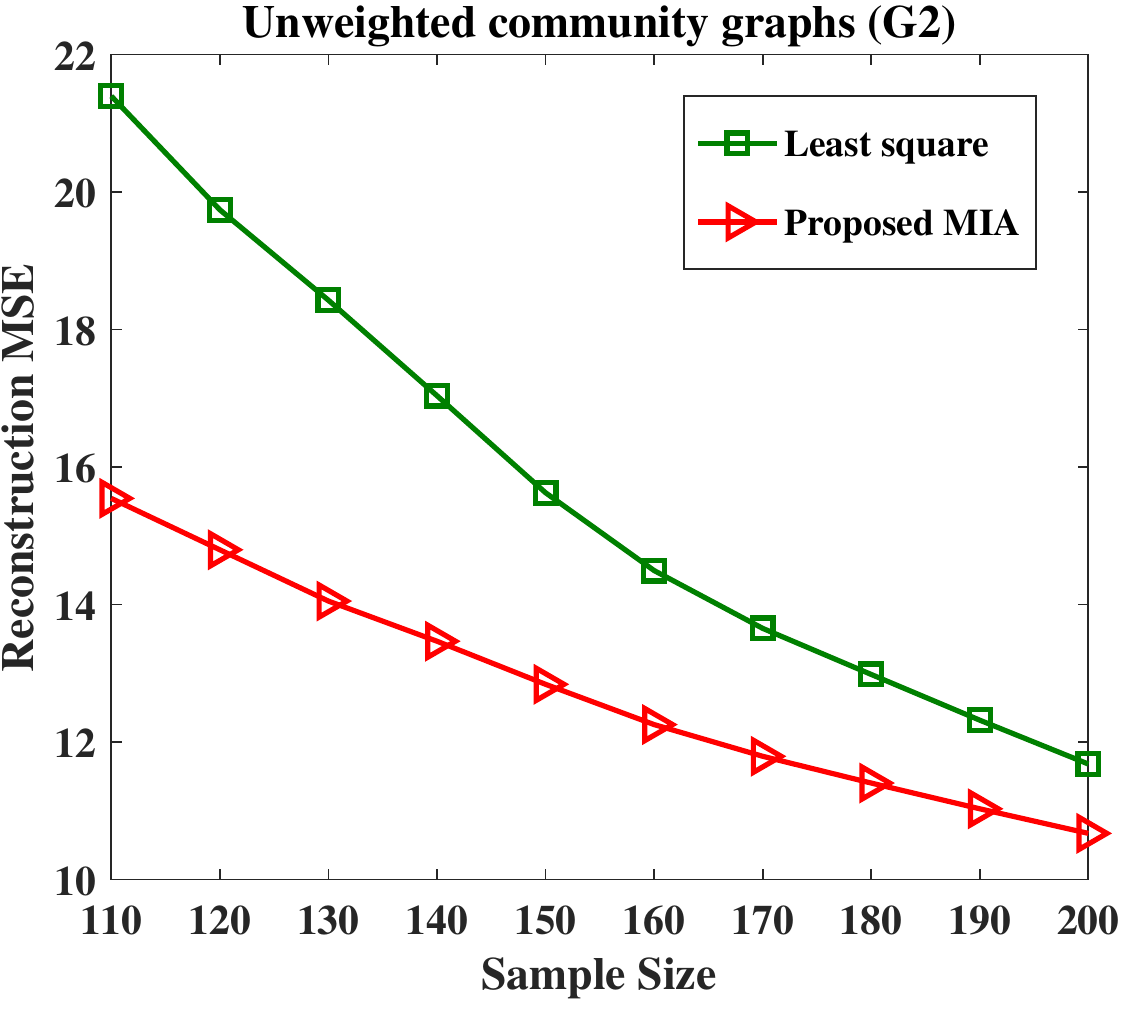}
\caption{Reconstruction MSE for different reconstruction algorithms in \textit{\textbf{G1}} and \textit{\textbf{G2}} at 0dB where the sampling algorithm are all the MIA sampling.}
\end{center}
\end{figure}

Combining \eqref{structure}, \eqref{Neumann recovery} and \eqref{series}, a closed-form reconstruction strategy (named as MIA reconstruction) is given by
\begin{align}\label{final recovery}
&\hspace{0cm}\hat{\mathbf{x}}= \mathbf{T} {\sum\limits_{l = 0}^\infty {{( {\mathbf{I} -\mathbf{P} } )}^l} }\mathbf{C}^{T} \mathbf{y}_\mathcal{S}\nonumber\\
&\hspace{0.2cm}=\mathbf{T}{\sum\limits_{l = 0}^\infty { \left[ \begin{array}{l}
 ({\mathbf{I}_{\mathcal{S}}} - {\mathbf{T}_\mathcal{S}})^{l}~~\bullet\\
~~~~~\mathbf{0} ~~~~~~~~~{\mathbf{I}_{{\mathcal{S}^c}}} \\
 \end{array} \right]\left[ \begin{array}{l}
\mathbf{y}_\mathcal{S}   \\
{\kern 1pt}{\kern 1pt}\mathbf{0} \\
 \end{array} \right]  } }\\
 &\hspace{0.2cm}=\mathbf{T}{\sum\limits_{l = 0}^\infty { \left[ \begin{array}{l}
 ({\mathbf{I}_{\mathcal{S}}} - {\mathbf{T}_\mathcal{S}})^{l}\mathbf{y}_\mathcal{S}\\
 ~~~~~~~~~~  \mathbf{0} \\
 \end{array} \right]  } }= \mathbf{T}_{\mathcal{VS}}{\mathbf{\Gamma}}\mathbf{y}_\mathcal{S}\approx \mathbf{T}^\textrm{Poly}_{\mathcal{VS}}\tilde{\mathbf{\Gamma}}\mathbf{y}_\mathcal{S},\nonumber
 \end{align}
where $ \mathbf{\Gamma}={\sum_{l = 0}^\infty {{{( {\mathbf{I}}_{\mathcal{S}} - {\mathbf{T}_\mathcal{S}} )}^l}} }$ and $ \tilde{\mathbf{\Gamma}}={\sum_{l = 0}^L {{{( {\mathbf{I}}_{\mathcal{S}} - {\mathbf{T}^{\textrm{Poly}}_\mathcal{S}} )}^l}} }$.

$\mathbf{T}^\textrm{Poly}$ and $\tilde{\mathbf{\Gamma}}$ have been computed in Table I, so the MIA recovery strategy only needs matrix-vector product, thus has low complexity.
Moreover, assuming the Chebyshev polynomial approximates the ideal low-pass filter well enough, this proposed MIA reconstruction method is more robust to large noise than the LS reconstruction in theory.
See Appendix B in the supporting document for the proof of the robustness.

\vspace{-0.05in}
\section{Experimental Results}

We evaluated our proposed strategy via simulations.
All experiments were performed in MATLAB R2017b, running on a PC with Intel Core I3 3.7 GHz CPU and 16GB RAM.

\emph{\textbf{Artificial graphs:}} \emph{\textbf{(G1)}} {Small-world} graphs \cite{smallWorld} (unweighted) with 1000 nodes, degree 8 and connection probability 0.1; \emph{\textbf{(G2)}}  Community graphs \cite{GSPtool} (unweighted) with 1000 nodes.

\emph{\textbf{Artificial signals}}: The true signal is exactly bandlimited with $K=50$ and the non-zero GFT coefficients are generated from $\mathcal{N}\left(1,0.5^{2}\right)$.
Samples are corrupted by AWGN.

\emph{\textbf{Other Parameters}}: We set $L=10$ for the MIA algorithm and $k=10$ for the spectral proxies algorithm.
{In the complexity comparison experiments, we set $R=10$, $T_1=100$ and $|\mathcal{E}|=\mathcal{O}(n)$.}
The SGWT toolbox \cite{GSPtool} is adopted to approximate the ideal low pass filter, where $p=25$ and $\alpha=30$ \cite{semisupervised}.

Fig.\,1(a), (b) and (c) show that our proposed MIA sampling algorithm achieves better MSE performance than the E-optimal and spectral proxies algorithms and {closely approximates the performance of the MFN algorithm in both {small-world} graphs and community graphs at different SNRs.}
{Fig.1 (d) shows that although the complexity of the MIA algorithm for the preparation step may be larger for a constant $m$, when $m$ is a fixed percentage of $n$, the proposed MIA algorithm has smaller complexity for both the preparation step and the sampling step compared to the MFN algorithm, especially for large graphs.}
To evaluate the Neumann truncation error at $L=10$, we computed the ratio between the estimate error in \eqref{trace error} and the MSE value in \eqref{original formulation} in {small-world} graphs.
Numerical results reveal {that} when $m=120$, this ratio was 0.19.

We also performed simulations using our proposed MIA reconstruction method, where the sample sets were all collected by the MIA sampling algorithm.
As depicted in Fig.\,2, the MIA reconstruction outperformed the LS reconstruction in both {small-world} graphs and community graphs at 0dB.
These results empirically validate the robustness of the proposed MIA reconstruction algorithm for large noise variance.

\section*{Acknowledgments}
The authors would like to thank Dr. Aamir Anis, Dr. Xuan Xue and the anonymous reviewers for constructive comments that led to improvements in the manuscript.

~~\\
~~\\
~~\\
~~\\
~~\\
\begin{center}
APPENDIX
\end{center}

A. Proof of Proposition 1

Assuming {that} the eigen-decomposition of $\mathbf{\Phi}$ is $\mathbf{\Phi}=\mathbf{U} \mathbf{\Lambda} {{\mathbf{U}^{T}}}$
where $\mathbf{\Lambda}=\textrm{diag}\{\delta _i\}$ and $\mathbf{U}\mathbf{U}^T=\mathbf{I}$.
Then,
\begin{align}
{\sum\limits_{l } {{{ {\bf{\Phi }} }^l}} }
= \sum\limits_{l } {\mathbf{U} \mathbf{\Lambda} {{\mathbf{U}^{T}}}...\mathbf{U} \mathbf{\Lambda} {{\mathbf{U}^{T}}}}
= \sum\limits_{l }{\mathbf{U} \mathbf{\Lambda}^{l} {{\mathbf{U}^{T}}}}
= {\bf{U}}{\mathbf{\Omega}}{{\bf{U}}^T}, \nonumber
\end{align}
where ${\mathbf{\Omega}}={\rm{diag}}\{{\sum\limits_{l}{{{ {\delta^l _i} }}} }\}$, $\delta_i$ is the $i$-th eigenvalue of $\mathbf{\Phi}$ and $0 \le {\delta _i}<1$ has been proved in Proposition 1.

Therefore,
\begin{align}\label{traceSum}
&\hspace{0cm}{{{\rm{tr}}\left( {\sum\limits_{l = L+1}^\infty  {{{ {\bf{\Phi }} }^l}} } \right)}
=\sum\limits_{i=1}^K {{\lambda _i}\left( {\sum\limits_{l = L+1}^\infty  {{{ {\bf{\Phi }} }^l}} } \right)}}\nonumber\\
&\hspace{0.3cm}=\sum\limits_{i=1}^K {\sum\limits_{l = L+1}^\infty  {\delta _i^l} }\\
&\hspace{0.3cm}= \sum\limits_{i = 1}^K {\frac{{  {\delta ^{L + 1}_i}  (1-{\delta ^{\infty}_i})   }}{{1 - {\delta _i}}}}\nonumber\\
&\hspace{0.3cm}=\sum\limits_{i = 1}^K {\frac{{{\delta ^{L + 1}_i}}}{{1 - {\delta _i}}}}>0,\nonumber
\end{align}
where ${\lambda _i}(\bullet)$ denotes the $i$-th eigenvalue of a matrix.

As a result,
\begin{align}
\left| {{\rm{tr}}\left[ {\sum\limits_{l = 0}^\infty  {{{ {\bf{\Phi }} }^l}} } \right] - {\rm{tr}}\left[ {\sum\limits_{l = 0}^L {{{ {\bf{\Phi }} }^l}} } \right]} \right|
=\left| {{\rm{tr}}\left[ {\sum\limits_{l = L+1}^\infty  {{{ {\bf{\Phi }} }^l}} } \right]} \right|
= \sum\limits_{i = 1}^K {\frac{{{\delta ^{L + 1}_i}}}{{1 - {\delta _i}}}}.\nonumber
\end{align}
\begin{flushright}
$\Box$
\end{flushright}

B. Proof of the robustness of the MIA reconstruction

{Assume that} the graph signal has the same energy for different SNR, i.e., ${\bf{E}}[{\bf{x}}{\bf{x}}^T]$ is a constant matrix, and the distribution of noise $\mathbf{n}$ is iid with zero mean and variance $\sigma^2$ which varies with SNR.
A corrupted $K$-BL graph signal is $\mathbf{y}_\mathcal{S}=\mathbf{x}_\mathcal{S}+\mathbf{n}$.

1) Least square (LS) reconstruction

If original signal is recovered by the LS reconstruction method, i.e.,
\begin{align}
&\hspace{0cm}{\bf{\hat x}_{\textrm{LS}}} = {{\bf{V}}_K}{({\bf{C}}{{\bf{V}}_K})^ {\dagger} }{{\bf{y}}_S}\nonumber\\
&\hspace{0.1cm}= {{\bf{V}}_K}{({\bf{C}}{{\bf{V}}_K})^ {\dagger} }({{\bf{x}}_S} + {\bf{n}})\nonumber\\
&\hspace{0.1cm}= {\bf{x}} + {{\bf{V}}_K}{({\bf{C}}{{\bf{V}}_K})^ {\dagger} }{\bf{n}},\nonumber
\end{align}
the {expected} mean square error (MSE) is
\begin{align}\label{LSexpectation}
&\hspace{0cm} {\bf{E}}\left\| {{\bf{\hat x}_{\textrm{LS}}} - {\bf{x}}} \right\|_2^2 = {\bf{E}}\left\| {{{\bf{V}}_K}{{({\bf{C}}{{\bf{V}}_K})}^ {\dagger} }{\bf{n}}} \right\|_2^2 \nonumber\\
&\hspace{0.3cm}= {\bf{E}}\left[ {{\rm{tr}}\left[ {\left( {{{\bf{V}}_K}{{({\bf{C}}{{\bf{V}}_K})}^ {\dagger} }{\bf{n}}} \right){{\left( {{{\bf{V}}_K}{{({\bf{C}}{{\bf{V}}_K})}^ {\dagger} }{\bf{n}}} \right)}^T}} \right]} \right]\nonumber\\
&\hspace{0.3cm}= {\rm{tr}}\left[ {{\bf{E}}\left( {{{\bf{V}}_K}{{({\bf{C}}{{\bf{V}}_K})}^ {\dagger} }{\bf{n}}{{\bf{n}}^T}{{({\bf{C}}{{\bf{V}}_K})}^ {\dagger} }{{\bf{V}}^T_K}} \right)} \right]\nonumber\\
&\hspace{0.3cm}= {\rm{tr}}\left[ {{{\bf{V}}_K}{{({\bf{C}}{{\bf{V}}_K})}^ {\dagger} }{\bf{E}}({\bf{n}}{{\bf{n}}^T}){{({\bf{C}}{{\bf{V}}_K})}^ {\dagger} }{{\bf{V}}^T_K}} \right]\\
&\hspace{0.3cm}= {\rm{tr}}\left[ {{{\bf{V}}_K}{{({\bf{C}}{{\bf{V}}_K})}^ {\dagger} }({\sigma ^2}{\bf{I}}){{({\bf{C}}{{\bf{V}}_K})}^ {\dagger} }{{\bf{V}}^T_K}} \right]\nonumber\\
&\hspace{0.3cm}= {\sigma ^2}{\rm{tr}}\left[ {{{\left( {{{({\bf{C}}{{\bf{V}}_K})}^T}({\bf{C}}{{\bf{V}}_K})} \right)}^{ - 1}}} \right].\nonumber
\end{align}

Moreover,
\begin{align}
&\hspace{0cm}{\rm{tr}}\left[ {{{\left( {{{({\bf{C}}{{\bf{V}}_K})}^T}({\bf{C}}{{\bf{V}}_K})} \right)}^{ - 1}}}\right]={{\rm{tr}}\left[ {\sum\limits_{l = 0}^\infty  {{{ {\bf{\Phi }} }^l}} } \right]}\nonumber\\
&\hspace{0.3cm}\mathop=\limits^{\bigtriangleup} \sum\limits_{i=1}^K {\sum\limits_{l = 0}^\infty  {\delta _i^l} }= \sum\limits_{i = 1}^K {\frac{1}{{1 - {\delta _i}}}},\nonumber
\end{align}
where the first equality holds because of Proposition 1 and $\mathop =\limits^{\bigtriangleup}$ holds by reusing the same derivation used in \eqref{traceSum}.

Therefore,
\begin{align}\label{LSMSE}
&\hspace{0cm} {\bf{E}}\left\| {{\bf{\hat x}_{\textrm{LS}}} - {\bf{x}}} \right\|_2^2
= {\sigma ^2}\sum\limits_{i = 1}^K {\frac{1}{{1 - {\delta _i}}}}.
\end{align}
2) Proposed matrix inversion approximation (MIA) reconstruction

According to Proposition 1 in our paper, the above LS reconstruction is equivalent to
\begin{align}
&\hspace{0cm}\hat{\mathbf{x}}_{\textrm{LS}}={\mathbf{V}_K}\left[ ({\mathbf{C}{\mathbf{V}_K})^{T}}\mathbf{C}{\mathbf{V}_K} \right]^{-1}({\mathbf{C}\mathbf{V}_K})^{T}\mathbf{y}_\mathcal{S}\nonumber\\
&\hspace{0.5cm}={\mathbf{V}_K} {\sum\limits_{l = 0}^\infty {{{\mathbf{\Phi}}^l}} } ({\mathbf{C}\mathbf{V}_K})^{T}\mathbf{y}_\mathcal{S}.
\nonumber
 \end{align}

If an original signal is recovered by the proposed MIA method, i.e., truncating the first $L$ items of the infinite matrix polynomial,
\begin{align}
&\hspace{0cm}{\bf{\hat x}_{\textrm{MIA}}} = {{\bf{V}}_K}\sum\limits_{l = 0}^L {{{\bf{\Phi }}^l}} {({\bf{C}}{{\bf{V}}_K})^{T}}({{\bf{x}}_S} + {\bf{n}})\nonumber\\
&\hspace{0cm}= {{\bf{V}}_K}\left[ {\sum\limits_{l = 0}^\infty  {{{\bf{\Phi }}^l}}  - \sum\limits_{l = L + 1}^\infty  {{{\bf{\Phi }}^l}} } \right]{({\bf{C}}{{\bf{V}}_K})^{T}}{{\bf{x}}_S} \nonumber\\
&\hspace{4.8cm}+ {{\bf{V}}_K}\sum\limits_{l = 0}^L {{{\bf{\Phi }}^l}} {({\bf{C}}{{\bf{V}}_K})^{T}}{\bf{n}}\nonumber\\
&\hspace{0cm}={\bf{x}} - {{\bf{V}}_K}\sum\limits_{l = L + 1}^\infty  {{{\bf{\Phi }}^l}} {({\bf{C}}{{\bf{V}}_K})^{T}}{{\bf{x}}_S} + {{\bf{V}}_K}\sum\limits_{l = 0}^L {{{\bf{\Phi }}^l}} {({\bf{C}}{{\bf{V}}_K})^{T}}{\bf{n}}\nonumber\\
&\hspace{0cm}:={\bf{x}} - \mathbf{t} + {{\bf{V}}_K}\sum\limits_{l = 0}^L {{{\bf{\Phi }}^l}} {({\bf{C}}{{\bf{V}}_K})^{T}}{\bf{n}},\nonumber
\end{align}
{where $\mathbf{t}$ is a vector representing the Von Neumann series truncation error on the bandlimited signal $\mathbf{x}_\mathcal{S}$ itself, that remains constant for different noise variance.}

Then, the corresponding MSE of the MIA reconstruction is
{\begin{align}\label{}
&\hspace{0cm}{\bf{E}}\left\| {{\bf{\hat x}_{\textrm{MIA}}} - {\bf{x}}} \right\|_2^2
= {\bf{E}}\left\| { - {\bf{t}} + {{\bf{V}}_K}\sum\limits_{l = 0}^L {{{\bf{\Phi }}^l}} {{({\bf{C}}{{\bf{V}}_K})}^{T}}{\bf{n}}} \right\|_2^2 \nonumber\\
&\hspace{0cm}=\left\| {\bf{t}} \right\|_2^2 - 2{\bf{E}}\left[ {{{\bf{t}}^{T}}{{\bf{V}}_K}\sum\limits_{l = 0}^L {{{\bf{\Phi }}^l}} {{({\bf{C}}{{\bf{V}}_K})}^{T}}{\bf{n}}} \right] \nonumber\\
&\hspace{3.1cm}+ {\bf{E}}\left\| {{{\bf{V}}_K}\sum\limits_{l = 0}^L {{{\bf{\Phi }}^l}} {{({\bf{C}}{{\bf{V}}_K})}^{T}}{\bf{n}}} \right\|_2^2\\
&\hspace{0cm}=\left\| {\bf{t}} \right\|_2^2 + {\bf{E}}\left\| {{{\bf{V}}_K}\sum\limits_{l = 0}^L {{{\bf{\Phi }}^l}} {{({\bf{C}}{{\bf{V}}_K})}^{T}}{\bf{n}}} \right\|_2^2. \nonumber
\end{align}}

Using the same derivation in \eqref{LSexpectation} and the property of trace operation, we have
\begin{align}\label{}
&\hspace{0cm}{\bf{E}}\left\| {{{\bf{V}}_K}\sum\limits_{l = 0}^L {{{\bf{\Phi }}^l}} {{({\bf{C}}{{\bf{V}}_K})}^{T}}{\bf{n}}} \right\|_2^2 \nonumber\\
&\hspace{0cm}={\rm{tr}}\left[ {{\bf{E}}\left( {{{\bf{V}}_K}\sum\limits_{l = 0}^L {{{\bf{\Phi }}^l}} {{({\bf{C}}{{\bf{V}}_K})}^{T}}{\bf{n}}{{\bf{n}}^{T}}{\bf{C}}{{\bf{V}}_K}\sum\limits_{l = 0}^L {{{\bf{\Phi }}^l}} {{\bf{V}}^T_K}} \right)} \right] \nonumber\\
&\hspace{0cm}={\rm{tr}}\left[ {{{\bf{V}}_K}\sum\limits_{l = 0}^L {{{\bf{\Phi }}^l}} {{({\bf{C}}{{\bf{V}}_K})}^{T}}{\bf{E}}({\bf{n}}{{\bf{n}}^{T}}){\bf{C}}{{\bf{V}}_K}\sum\limits_{l = 0}^L {{{\bf{\Phi }}^l}} {{\bf{V}}^T_K}} \right] \nonumber\\
&\hspace{0cm}={\sigma ^2}{\rm{tr}}\left[ {\sum\limits_{l = 0}^L {{{\bf{\Phi }}^l}} {{({\bf{C}}{{\bf{V}}_K})}^{T}}{\bf{C}}{{\bf{V}}_K}\sum\limits_{l = 0}^L {{{\bf{\Phi }}^l}} } \right]\nonumber\\
&\hspace{0cm}={\sigma ^2}{\rm{tr}}\left[ {\left( {{\bf{I}} - {\bf{\Phi }}} \right){{\left( {\sum\limits_{l = 0}^L {{{ {\bf{\Phi }} }^l}} } \right)}^2}} \right]. \nonumber
\end{align}

Therefore,
\begin{align}
{\bf{E}}\left\| {{\bf{\hat x}_{\textrm{MIA}}} - {\bf{x}}} \right\|_2^2
 = \left\| {\bf{t}} \right\|_2^2 + {\sigma ^2}{\rm{tr}}\left[ {\left( {{\bf{I}} - {\bf{\Phi }}} \right){{\left( {\sum\limits_{l = 0}^L {{{\bf{\Phi }}^l}} } \right)}^2}} \right].\nonumber
 \end{align}

The \emph{Von Neumann's trace inequality} states that
\begin{align}
{\rm{tr}}\left[ {\left( {{\bf{I}} - {\bf{\Phi }}} \right){{\left( {\sum\limits_{l = 0}^L {{{\bf{\Phi }}^l}} } \right)}^2}} \right] \le \sum\limits_{i = 1}^K {{\lambda _i}\left( {{\bf{I}} - {\bf{\Phi }}} \right){\lambda _i}} \left[ {{{\left( {\sum\limits_{l = 0}^L {{{\bf{\Phi }}^l}} } \right)}^2}} \right],\nonumber
 \end{align}
where ${\lambda _i}(\bullet)$ denotes the $i$-th eigenvalue of a matrix.

By reusing the derivation in \eqref{traceSum}, we know
\begin{align}
{\lambda _i} \left[ {{{\left( {\sum\limits_{l = 0}^L {{{\bf{\Phi }}^l}} } \right)}^2}} \right]
={\lambda^2 _i}  {{{\left( {\sum\limits_{l = 0}^L {{{\bf{\Phi }}^l}} } \right)}}}
=\left(\sum\limits_{l = 0}^L {{\delta^l _i}}\right)^2=\frac{{(1 - {\delta ^{L + 1}_i})^2}}{{(1 - {\delta _i})}^2}.\nonumber
 \end{align}

Combined with ${\lambda _i}\left( {{\bf{I}} - {\bf{\Phi }}} \right)=1-{\delta _i}$,
\begin{align}
{\rm{tr}}\left[ {\left( {{\bf{I}} - {\bf{\Phi }}} \right){{\left( {\sum\limits_{l = 0}^L {{{\bf{\Phi }}^l}} } \right)}^2}} \right] \le \sum\limits_{i = 1}^K \frac{{(1 - {\delta ^{L + 1}_i})^2}}{{1 - {\delta _i}}},
 \end{align}
which results in
\begin{align}\label{MIAmse}
{\bf{E}}\left\| {{\bf{\hat x}_{\textrm{MIA}}} - {\bf{x}}} \right\|_2^2
  \le \left\| {\bf{t}} \right\|_2^2 + {\sigma ^2}\sum\limits_{i = 1}^K \frac{{(1 - {\delta ^{L + 1}_i})^2}}{{1 - {\delta _i}}}.
 \end{align}

We have proved in Proposition 1 that $0 \le {\delta _i}<1$, so
\begin{align}\label{inequality}
\sum\limits_{i = 1}^K \frac{{(1 - {\delta ^{L + 1}_i})^2}}{{1 - {\delta _i}}}<\sum\limits_{i = 1}^K {\frac{1}{{1 - {\delta _i}}}}.
 \end{align}

{Combining \eqref{LSMSE}, \eqref{MIAmse} and \eqref{inequality}}, we can see that when noise variance, i.e., $\sigma^2$, is very large, the proposed MIA reconstruction method will achieve better MSE performance.
Thus, we can safely claim that our proposed reconstruction method is more robust to large noise than the LS reconstruction.
\hfill $\Box$


\begin{thebibliography}{}

\bibitem{emerging}
D. Shuman, S. Narang, P. Frossard, A. Ortega, and P. Vandergheynst, ``The emerging field of signal processing on graphs: Extending high-dimensional data analysis to networks and other irregular domains,'' {\emph{IEEE Signal Process. Mag.}}, vol. 30, no. 3, pp. 83-98, May. 2013.

\bibitem{gsp}
A. Sandryhaila and J. Moura, ``Big data analysis with signal processing
on graphs: Representation and processing of massive data sets with irregular structure,'' {\emph{IEEE Signal Process. Mag.}}, vol. 31, no. 5, pp. 80-90,
Sep. 2014.

\bibitem{sampling}
I. Pesenson, ``Sampling in Paley-Wiener spaces on combinatorial
graphs,'' {\emph{Trans. Amer. Math. Soc.}}, vol. 360, no. 10, pp. 5603-5627,
2008.

\bibitem{randomSampling}
G. Puy, N. Tremblay, R. Gribonval, and P. Vandergheynst, ``Random sampling of bandlimited signals on graphs,''  {\emph{Appl. Comput. Harmon. Anal.}}, 2016.

\bibitem{Uncertainty}
M. Tsitsvero, S. Barbarossa, and P. Di Lorenzo, {``Signals on graphs: Uncertainty principle and sampling,'' } {{\emph{IEEE Trans. Signal Process.}}}, vol. 64, no. 18, pp. 4845-4860, Sept.15, 2016.


\bibitem{sketching}
F. Gama, A. G. Marques, G. Mateos, and A. Ribeiro, {``Rethinking sketching as sampling:
A graph signal processing approach,'' }\textit{arXiv preprint }, arXiv:1611.00119, 2016.

\bibitem{aggregation}
{A. Marques, S. Segarra, G. Leus, and A. Ribeiro,``Sampling of graph
signals with successive local aggregations,'' {\emph{IEEE Trans. Signal Process.}},
vol. 64, no. 7, pp. 1832-1843, 2016.}

\bibitem{largedata}
I. Shomorony and A. Avestimehr, ``Sampling large data on graphs,'' in
{\emph{Proc. IEEE Global Conf. Signal Inf. Process. (GlobalSIP)}}, pp. 933-936, Dec. 2014.

\bibitem{towards}
A. Anis, A. Gadde, and A. Ortega, ``Towards a sampling theorem
for signals on arbitrary graphs,'' in {\emph{Proc. IEEE Int. Conf. Acoust.,
Speech, Signal Process. (ICASSP)}}, Florence, Italy, pp. 3864-3868,
May. 2014.

\bibitem{semisupervised}
A. Gadde, A. Anis, and A. Ortega, ``Active semi-supervised learning using sampling theory for graph signals,'' in \emph{Proc. 20th ACM SIGKDD Int. Conf.
Knowl. Discov. Data Min.}, pp. 492-501, 2014.

\bibitem{AO}
A. Anis, A. Gadde, and A. Ortega, ``Efficient sampling set selection for
bandlimited graph signals using graph spectral proxies,'' {\emph{IEEE Trans.
Signal Process.}}, vol. 64, no. 14, pp. 3775-3789, Jul. 2016.

\bibitem{SCsampling}
S. Chen, R. Varma, A. Sandryhaila, and {J. Kova{\v{c}}evi{\'{c}}}, ``Discrete signal
processing on graphs: Sampling theory,'' {\emph{IEEE Trans. Signal Process.}},
vol. 63, no. 24, pp. 6510-6523, 2015.

\bibitem{Boyd}
S. Boyd and L. Vandenberghe, {\emph{Convex Optimization}}, Cambridge University Press, 2004.

\bibitem{greedybound}
L. F. O. Chamon and A. Ribeiro, {``Greedy sampling of graph signals,''} {{\emph{IEEE Trans. Signal Process.}}}, vol. 66, no. 1, pp. 34-47, Jan.1, 2018.

\bibitem{wavelet}
D. Hammond, P. Vandergheynst, and R. Gribonval, ``Wavelets on graphs via spectral graph theory,'' {\emph{Appl. Comput. Harmon. Anal.}}, vol. 30, no. 2, pp. 129-150, 2011.

 \bibitem{codeguide}
S. Chen, A. Sandryhaila, and {J. Kova{\v{c}}evi{\'{c}}}, ``Sampling theory for graph signals,'' in {\emph{Proc. IEEE Int. Conf. Acoust., Speech, Signal Process. (ICASSP)}}, South Brisbane, Queensland, Australia, pp. 3392-3396,
Apr. 2015.

\bibitem{experiments}
F. Pukelsheim, {\emph{Optimal Design of Experiments}}. Philadelphia, PA, USA: SIAM, 1993, vol. 50.

\bibitem{neumannseries}
R. A. Horn and C. R. Johnson, \emph{Matrix Analysis.} Cambridge University Press, 2013.

\bibitem{ILSR}
S. K. Narang, A. Gadde, E. Sanou, and A. Ortega, ``Localized iterative methods for interpolation in graph structured data,'' in {\emph{Proc. IEEE Global Conf. Signal Inf. Process. (GlobalSIP)}}, Austin,USA, pp. 491-494, 2013.

\bibitem{golub}
G. Golub and C. F. V. Loan, \textit{Matrix Computations} (Johns Hopkins Studies in the Mathematical Sciences). Johns Hopkins University Press, 2012.

\bibitem{Coakley}
E. S. Coakley and V. Rokhlin, ``A fast divide-and-conquer algorithm for computing the spectra of real symmetric tridiagonal matrices,'' {\emph{Appl. Comput. Harmon. Anal.}}, 2013.

\bibitem{matrix}
{V. V. Williams}, ``Multiplying matrices in $\mathcal{O}(n^{2.373})$ time,'' Stanford University, July 1, 2014.

\bibitem{smallWorld}
D. J. Watts and S. H. Strogatz, ``Collective dynamics of small-world networks,'' \emph{Nature}, vol. 393, no. 6684, pp. 440-442,
1998.

\bibitem{GSPtool}
N. Perraudin, J. Paratte, D. Shuman, V. Kalofolias, P. Vandergheynst, and D. K. Hammond, ``GSPBOX: A toolbox for signal processing on graphs,''
Aug. 2014, arXiv:1408.5781 [cs.IT].
\end{thebibliography}
\end{document}